\documentclass[%
 reprint,
 amsmath,
 amssymb,
 showkeys,
 aip,
 jcp,
]{revtex4-2}

\usepackage{graphicx}
\usepackage{dcolumn}
\usepackage{bm}
\usepackage{xcolor}
\usepackage[T1]{fontenc}
\usepackage[normalem]{ulem}
\usepackage{amssymb}

\newcommand{\rhoeq}{\rho^\mathrm{eq}}
\newcommand{\taud}{\tau_\mathrm{D}}

\newcommand{\tauc}{\tau_\mathrm{c}}
\newcommand{\kT}{k_\mathrm{B}T}

\begin{document}

\title{Simultaneous optimization of assembly time and yield in programmable self-assembly}

\author{Maximilian C. H\"ubl}
\email{maximilian.huebl@ist.ac.at}
\author{Carl P. Goodrich}%
\email{carl.goodrich@ist.ac.at}
\affiliation{%
 Institute of Science and Technology Austria (ISTA), Am Campus 1, 3400 Klosterneuburg, Austria
}%


\begin{abstract}
Rational design strategies for self-assembly require a detailed understanding of both the equilibrium state and the assembly kinetics. While the former is starting to be well understood, the latter remains a major theoretical challenge, especially in programmable systems and the so-called semiaddressable regime, where binding is often nondeterministic and the formation of off-target structures negatively influences the assembly. Here, we show that it is possible to simultaneously sculpt the assembly outcome and the assembly kinetics through the underexplored design space of binding energies and particle concentrations. 
By formulating the assembly process as a complex reaction network, we calculate and optimize the tradeoff between assembly speed and quality, and show that parameter optimization can speed up assembly by many orders of magnitude without lowering the yield of the target structure. Although the exact speedup varies from design to design, we find the largest speedups for nondeterministic systems where unoptimized assembly is the slowest, sometimes even making them assemble faster than optimized fully-addressable designs.
Therefore, these results not only solve a key challenge in semiaddressable self-assembly, but further emphasize the utility of semiaddressability, where designs have the potential to be faster as well as cheaper (fewer particle species) and better (higher yield).
More broadly, our results highlight the importance of parameter optimization in programmable self-assembly, and provide practical tools for simultaneous optimization of kinetics and yield in a wide range of systems.
\end{abstract}

\maketitle

\section{Introduction}
Predicting and controlling self-assembly processes is a fundamental challenge with major implications throughout nanotechnology, medicine, and biology~\cite{Whitesides.2002, Grzybowski.2009, Cademartiri.2015, Yadav.2020, Hagan.2021, Israelachvili2011}.
A common and powerful approach is to design particle-particle interactions so that the desired assembly outcome is the thermodynamic ground state, meaning that as the system equilibrates, this desired outcome naturally emerges. 
The difficulty with this approach is that many self-assembling systems do not equilibrate on experimentally accessible timescales, making understanding and optimizing assembly kinetics one of the key challenges in self-assembly. 

Thermodynamic ground states can be engineered in a number of ways. For example, many experimental platforms dictate the size and shape of constituent particles while also programming specific binding rules that govern where bonds are allowed to form, see Fig.~1(a). These binding rules, with either deterministic or nondeterministic interactions, are extremely important in prescribing both what structures are capable of assembling in principle and which will assemble with high equilibrium yield~\cite{Huebl.2025, Bohlin.2023, Hayakawa.2022pqs, Hayakawa.2024, Videbaek.2024}. Furthermore, for a given set of binding rules, there is a secondary design space composed of binding energies and particle concentrations that can be used to further refine, or even dramatically alter, equilibrium yield~\cite{Huebl.2025b, Murugan.2015}. However, this growing ability to precisely design equilibrium assembly outcomes has not addressed the challenge of kinetics. In fact, assembly time and assembly quality are often in competition with each other~\cite{Trubiano.2021, Jhaveri.2024, Huebl.2025m3}, meaning that the progress made in equilibrium assembly design is severely held back unless and until kinetics can be incorporated into the design process in a robust way.

\begin{figure*}
    \centering
    \includegraphics[width=\textwidth]{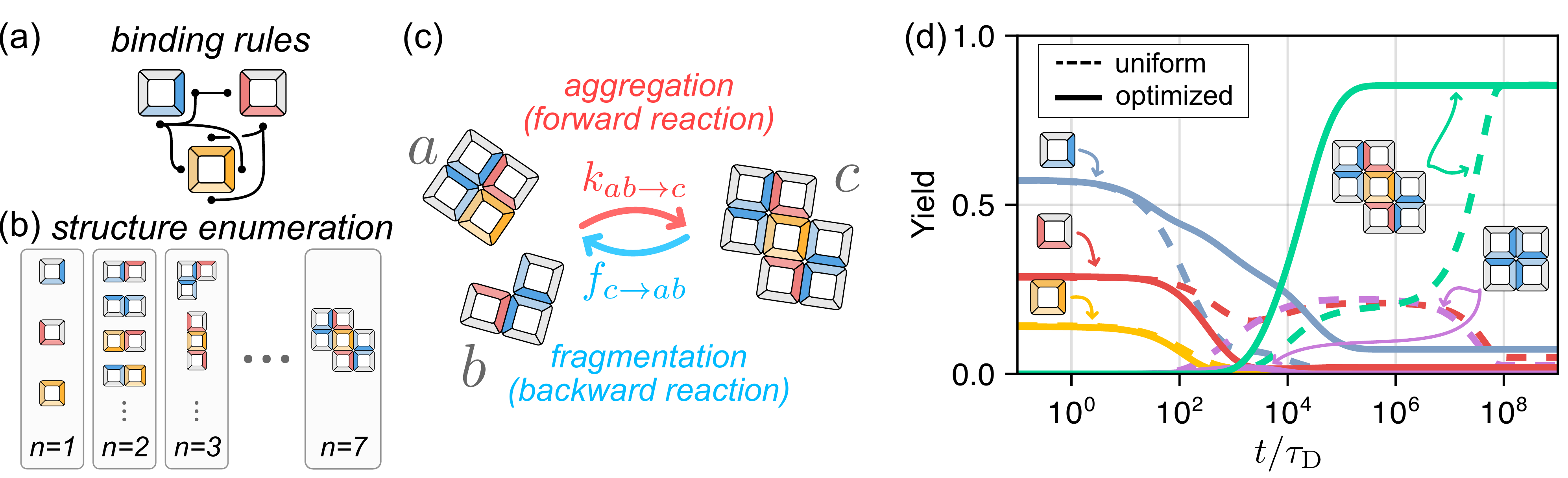}
    \caption{Assembly kinetics as a complex reaction network. (a) Binding rules showing selectivity of binding. Only particle sides that are connected by edges may bind to each other, thereby limiting what structures may form. (b) From the binding rules, we enumerate all possible structures, as illustrated here by sketching all possible monomers, dimers, trimers, and so on. For the rules shown in (a), only 71 structures are possible, the largest of which is shown on the right. (c) To investigate the assembly kinetics, we construct all possible reactions between all structures. Shown here is one of 268 possible reaction pairs resulting from the binding rules in (a). The rate constant for two structures $a$ and $b$ to combine into a structure $c$ is given by $k_{ab\to c}$, whereas the fragmentation of $c$ into $a$ and $b$ is controlled by the rate $f_{c \to ab}$. (d) Time evolution of the assembly process for uniform binding energies fixed at $12 \, \kT$ (dashed lines), and optimized unequal binding energies (solid lines), as discussed in the text. Shown are the time dependent yields of three monomers (blue, red, yellow), an off-target four-particle square (purple), and the seven-particle target structure (teal). Total particle concentration is $0.01/\sigma^3$. Time is measured in units of the monomer diffusion time, $\tau_\mathrm{D}$.}
    \label{fig1}
\end{figure*}

Recent work on assembly kinetics has shed light on the effects of particle geometry~\cite{Gartner.2024} and target structure size~\cite{Gartner.2022}, and has highlighted different mechanisms for kinetic arrest along with design principles and assembly protocols for avoiding them~\cite{Murugan.2015, Jhaveri.2024, Whitelam.20158t, Hagan.2011}.
Moreover, it has recently been shown that there is a high amount of degeneracy in the design space of equilibrium programmable assembly~\cite{Huebl.2025b}, meaning that equivalent assembly outcomes can be achieved with a large range of different design parameters.
Exploiting this degeneracy may make it possible to optimize kinetics while leaving the equilibrium properties untouched, and without the need for complex, time-dependent assembly protocols.
But to perform this optimization in practice, we require an accurate, fast, and general framework that can relate the attributes of the assembling particles to the kinetics and equilibrium outcome of the self-assembling process.

Here, we introduce and apply a framework for predicting and optimizing assembly times alongside assembly yield.
By formulating the assembly process as a complex reaction network, we calculate the number density of each structure over time, similar to the approach of Jhaveri et al.~\cite{Jhaveri.2024}. 
However, because we derive the underlying reaction rates from the binding energies and chemical potentials, we can optimize the kinetics through the same design space used to optimize the equilibrium yield, allowing for simultaneous optimization. Furthermore, by exploiting the structure enumeration tools of Ref.~\cite{Huebl.2025} to identify possible reactions, we apply this deep into the semiaddressable regime, where particle types are reused and binding is nondeterministic. 

Through this, we often attain drastic assembly speedups, sometimes of many orders of magnitude, without having to compromise on assembly yield.
Internally, this is achieved by mitigating or outright avoiding kinetic traps caused by unwanted off-target structures, while also sculpting kinetic pathways to further minimize equilibration times. The benefits of our approach are most striking in multifarious, economical, or otherwise semiaddressable assembly designs, which after optimization often assemble faster and at higher yield compared to their fully-addressable counterparts where every particle in the desired structure is designed separately.
These results demonstrate a practical, fast, and ready-to-use method for predicting and designing both the assembly kinetics and equilibrium properties in a wide range of systems, from protein assembly to DNA nanotechnology.

\section{The assembly process as a complex reaction network}
Given a set of building blocks, each with multiple specific binding sites, we consider a fixed set of binding rules stipulating which pairs of binding sites are complementary and may bind (Fig.~\ref{fig1}(a)).
Imposing these discrete binding rules then allows us to carry out an enumeration of the possible resulting structures (Fig.~\ref{fig1}(b)).
We focus throughout on binding rules that are selective enough to lead to a finite set of possible structures, and we use the Roly.jl software~\cite{Huebl.2025} to enumerate all structures that are compatible with the binding rules.
From the structure enumeration, we can compute the equilibrium properties of the system, most notably the equilibrium number density and yield of each structure.
However, to gain insight into the assembly kinetics, we need an additional ingredient: the reactions between structures.

We model the assembly process as a complex reaction network, where structures can dynamically aggregate and fragment.
Given the enumeration of structures, we identify a list of reactions, consisting of all the ways the structures can break into two pieces, and all the ways two structures can combine to form a larger structure (Fig.~\ref{fig1}(c)).
Each pair of forward and backward reactions $r$ consists of a triple of structures, $a_r + b_r \rightleftharpoons c_r$, indicating that structures $a_r$ and $b_r$ can combine to form structure $c_r$, and that $c_r$ can fracture into $a_r$ and $b_r$.
Once all reactions and their corresponding reaction rates are identified, we can write down a system of ordinary differential equations that describes the time evolution of the system.
Denoting the density of a structure $i$ by $\rho_i(t)$, we have
\begin{align}\label{eq:ode}
    \frac{d\rho_i(t)}{dt} = \sum_{r \in \mathcal{R}} &[\delta_{i c_r} - (\delta_{i a_r} + \delta_{i b_r})] \, \times \\
    &[k_{a_r b_r \to c_r} \rho_{a_r} \rho_{b_r} - f_{c_r \to a_r b_r} \rho_{c_r}] \,, \nonumber
\end{align}
where $\mathcal{R}$ is the set of all reactions, $\delta_{ij}$ is the Kronecker delta, $k_{ab \to c}$ is the aggregation rate constant for the formation of structure $c$ from $a$ and $b$, and $f_{c \to ab}$ is the fragmentation rate constant of $c$ into $a$ and $b$.

Making this formulation of the assembly process meaningful requires reasonable expressions for the aggregation and fragmentation rates, $k_{ab \to c}$ and $f_{c \to ab}$.
This is a major theoretical challenge~\cite{Zhou.1993,  Solc.1971, Solc.1973, Berg.1985, Haenggi.1990, Schreiber.2009}.
While the diffusion-limited aggregation rate for isotropic spherical particles is given by the well-known Smoluchowski expression~\cite{Smoluchowski.1918}, the situation for anisotropic particles is much more complicated because their binding requires orientational alignment.
Ignoring this orientational alignment can lead to drastic overestimation of the aggregation rates~\cite{Solc.1971, Solc.1973, Berg.1985, Zhou.1993, Schreiber.2009}, resulting in assembly times that are wrong by orders of magnitude.

To obtain aggregation rates that qualitatively take orientational alignment into account, we adopt a coarse-grained description of the assembled structure's diffusive behavior.
Following previous work on protein aggregation~\cite{Zhou.1993, Schreiber.2009}, we model two aggregating structures as spheres with axially symmetric patches, where contact between the patches indicates the formation of the ``reactive complex'', a short-lived intermediate state that rapidly relaxes to the stable bound state.
For this spherical geometry, binding rates can be computed analytically and take the form
\begin{equation}\label{eq:rate}
    k_{ab \to c} = 4\pi (D_a + D_b) (R_a + R_b) \, \gamma_{ab \to c} \,,
\end{equation}
where $D_i$ is the translational diffusion coefficient of structure $i$, $R_i$ is the radius of structure $i$, and $\gamma_{ab\to c}$ is the orientational correction factor that determines the reduction of the aggregation rate compared to the Smoluchowski expression.
This factor depends on the translational and rotational diffusion constants of the particles, and on the area of the interaction patch.
In the limit of fast rotational diffusion and/or large patch area, $\gamma_{ab\to c} \to 1$ and Eq.~\eqref{eq:rate} converges to the Smoluchowski result for isotropic particles.
We calculate the diffusion constants of a structure by modeling it as a rigid sphere-array~\cite{Harvey.1980, Carrasco.1999, Kraft.2013}, and we estimate the fractional patch area by considering the distance between a structure's binding sites and its center of diffusion~\cite{Harvey.1980}.
All details regarding these calculations can be found in the SI.

Once all aggregation rates $k_{ab \to c}$ are known, the fragmentation rates $f_{c \to ab}$ can be determined by requiring detailed balance to hold~\cite{Vigil.2009, Kampen1992}.
If the equilibrium concentrations of structures are given by $\rhoeq_i$, we must have
\begin{equation}
    k_{ab \to c}\,\rhoeq_a \rhoeq_b = f_{c \to ab} \,\rhoeq_c \,,
\end{equation}
for the steady state of the reaction network to coincide with the imposed equilibrium concentrations~\footnote{Note that it would be, in principle, possible to compute the rates by going the other direction; i.e. compute the fragmentation rates first using, for example, Kramers theory, and then use detailed balance to compute the aggregation rates.
However, this method of computing the rates is severely complicated by the orientational degrees of freedom of the assembling particles, and the fact that there is not necessarily a unique saddle point along the reaction path, making our approach easier to apply in comparison.}.

Since the equilibrium concentrations can be calculated analytically using statistical mechanics~\cite{Klein.2018, curatolo.2023, Huebl.2025, Huebl.2025b}, the fragmentation rates are obtained straightforwardly.
Importantly, the fragmentation rates are controlled by the bond energies of the bonds that break when a structure $c$ fragments into $a$ and $b$:
\begin{equation}
    f_{c \to ab} \propto e^{(E_a + E_b - E_c) / k_\mathrm{B}T} \,,
\end{equation}
where the $E_i$ is the total bond energy in structure $i$.
This means that tuning the binding energies between building block particles allows us to tune the fragmentation rates $f_{c \to ab}$.
Besides tuning particle concentrations, this is the main way assembly kinetics can be controlled in our approach.

\section{Optimized design parameters speed up assembly by orders of magnitude}
\subsection{Uniform binding energies lead to kinetic traps}
With these tools in hand, we are ready to investigate the kinetics of various self-assembly processes.
We begin with the binding rules shown in Fig.~\ref{fig1}(a)~\footnote{The equilibrium properties of these binding rules were originally investigated in Ref.~\cite{Huebl.2025}}, which lead to 71 possible structures, the largest of which is the seven-particle shape shown on the right in both Fig.~\ref{fig1}(b) and (c).
Note that even though all structures are planar, we assume that they assemble in three dimensions when calculating the rate constants, which is also the case in many experimental systems.

To favor assembly of the seven-particle structure, we begin by making a simple and sensible choice for the assembly parameters: we choose particle concentrations to follow the stoichiometry of the target, and we set the binding energies of all six possible bonds equal to $12\, k_\mathrm{B}T$.
It was shown in Ref.~\cite{Huebl.2025} that this leads to high equilibrium yield, even though there exist various misbound off-target structures that are also compatible with the binding rules in Fig.~\ref{fig1}(a).
The resulting equilibration trajectory for these ``uniform'' parameters is shown as the dashed curves in Fig.~\ref{fig1}(d).
We choose the reference unit of time to be the characteristic diffusion time of the particles, $\tau_\mathrm{D} = \sigma^2 / D_0$, where $\sigma$ is the particle length scale and $D_0$ is the (translational) particle diffusion constant.

Even though the steady state target yield is high, the trajectory clearly shows signs of kinetic trapping: starting at around $10^5 \,\taud$, the yield of the target structure (teal dashed curve) starts stagnating because all blue monomers are used up by the 4-particle square, an off-target structure acting as a kinetic trap (purple dashed curve).
Equilibration is significantly delayed because the relatively high binding energies stabilize this 4-particle square, meaning that it takes a long time for the blue monomers trapped in the off-target structure to break free.
Due to this, the final relaxation to equilibrium only occurs around $t\approx10^8\,\taud$.

This highlights a general feature often observed in self-assembly: the tradeoff between equilibrium yield and equilibration time.
Reaching high equilibrium yield requires high binding energies to stabilize the target structures, but these interactions often also stabilize unwanted off-target structures, leading to kinetic arrest.

\begin{figure*}
    \centering
    \includegraphics[width=0.8\textwidth]{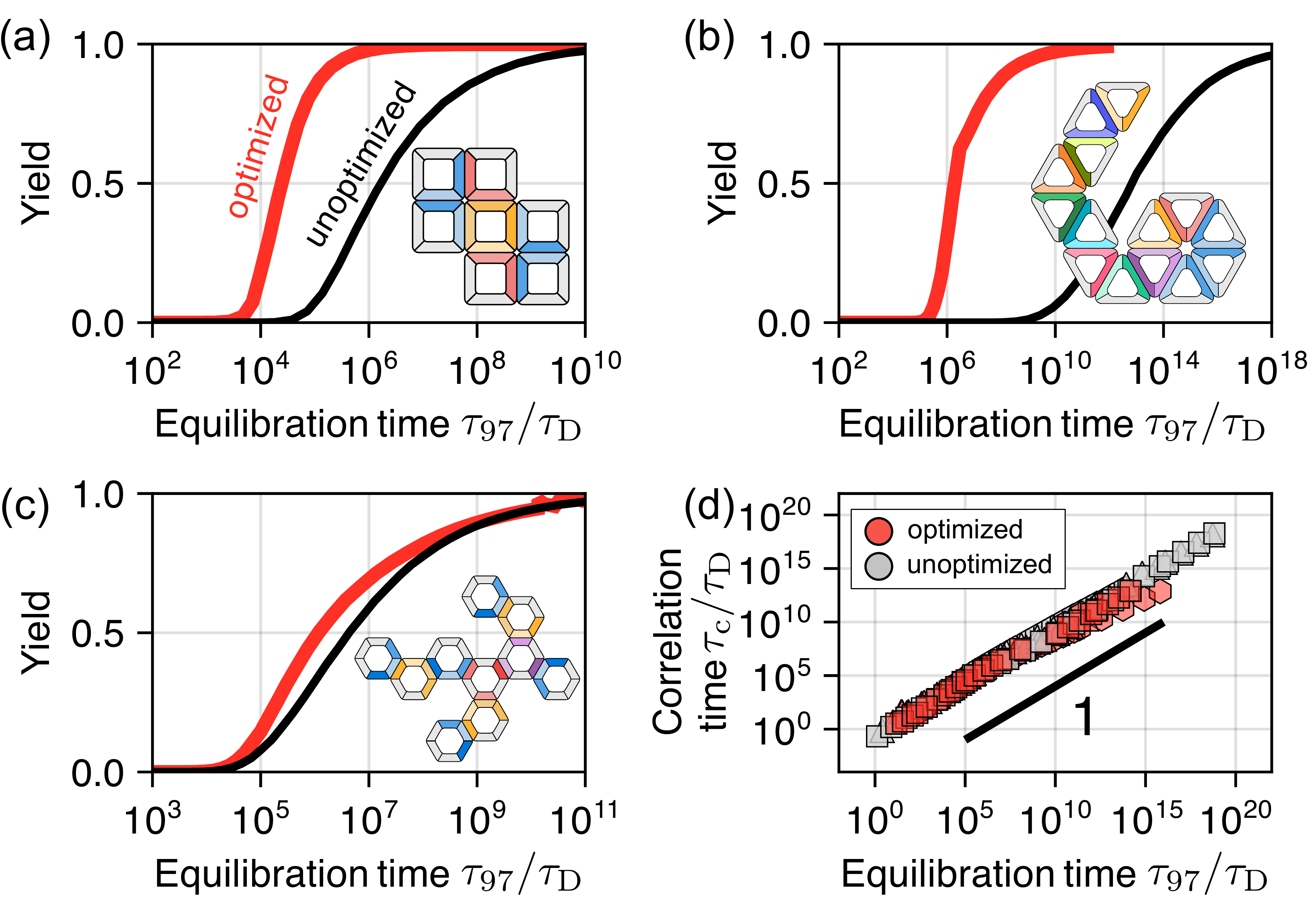}
    \caption{Yield and equilibration time tradeoff. Equilibrium yield as a function of equilibration time, for the seven-particle square shape (a), a 14-particle triangle shape (b), and a ten-particle hexagon shape (c). The different curves correspond to different methods of parameter optimization. Black solid curve: unoptimized, uniform binding energies and stoichiometric particle concentrations. Red solid curve: optimized parameters obtained by minimizing the correlation time $\tauc$ through the procedure described in Section~\ref{sec:minimize}, while keeping total particle concentration fixed at $0.01 / \sigma^3$. (d) Relationship between equilibration time (here defined to be the time $\tau_{97}$ where target yield first reaches $97\%$ of the equilibrium value) and correlation time $\tauc$ for different methods of parameter optimization (colors). Marker shape corresponds to the system: squares correspond to (a), triangles correspond to (b), and hexagons to (c). The black line has slope 1 and is shown for comparison.}
    \label{fig2}
\end{figure*}

\subsection{A fast minimization procedure for equilibration times}\label{sec:minimize}
To what extent is the kinetic trapping caused by our naive choice of design parameters (stoichiometric particle concentrations and uniform binding energies)?
Can we optimize our design parameters to avoid the kinetic trap caused by the four-particle square, while maintaining the same target yield?
To answer these questions, we now introduce a numerical optimization procedure for minimizing equilibration time.

The basic idea of this procedure is to pick some initial parameters (for example, uniform binding energies and stoichiometric particle concentrations, as discussed above), and then optimize these parameters to reduce the equilibration time while keeping the final target yield and the total particle concentration the same.
In this way, we can directly compare the results of the optimization to our initial, naive set of parameters.
By picking a series of initial guesses that cover target yields from $0\%$ to $100\%$, for example by scaling the initial binding energy up and down, we can then construct a yield-time tradeoff curve that shows the optimal balance between target yield and equilibration time.

While it is possible to ``brute-force'' optimize the parameters by calculating the equilibration time and its gradient with respect to the design parameters from the full solution of Eq.~\eqref{eq:ode}~\cite{Jhaveri.2024}, we find that it is much faster to exploit the relationship between the equilibration time and the \emph{correlation} time of the system.
Since the equilibration and correlation time of a system are generally positively correlated with each other, it is reasonable to expect that minimizing one will minimize the other (this will be verified \emph{a posteriori} below).
The correlation time, however, can be calculated much faster simply by a linear stability analysis around the equilibrium state.
Importantly, we already know the equilibrium state from statistical mechanics, and therefore do not need to solve the dynamical system, Eq.~\eqref{eq:ode}, at all during the optimization process.

To compute the correlation time, we linearize Eq.~\eqref{eq:ode} around the equilibrium state, so that for perturbations $\delta\rho_i$ around equilibrium, we have
\begin{equation}
    \frac{d\delta\rho_i}{dt} = \sum_j S_{ij} \,\delta\rho_j \,,
\end{equation}
where $S$ is the linear stability matrix, which depends on particle chemical potentials and binding energies, and whose full expression is provided in SI.
The eigenvalues of $S$ are related to the decay times of their corresponding eigenmodes.
For a system with $n_\mathrm{p}$ particle species, $S$ in general has $n_\mathrm{p}$ zero-modes, which correspond to changes in the particle concentrations.
Since changing particle concentrations changes the equilibrium state, these modes do not decay.
All other eigenvalues of $S$ must have negative real parts. 
From the non-zero eigenvalue with the largest real part, $\lambda_{n_\mathrm{p} + 1}$, we can define the correlation time as $\tauc = - \mathrm{Re}(\lambda_{n_\mathrm{p} + 1})^{-1}$.
As an interesting side note, we consistently find that all eigenvalues of $S$ are purely real even though $S$ is not symmetric, suggesting that there exists some internal symmetry that makes $S$ quasi-hermitian.

The linearization procedure described above gives $\tauc(\bm{\mu}, \bm{\varepsilon})$ as a function of particle chemical potentials $\bm{\mu}$ and binding energies $\bm{\varepsilon}$.
$\tauc$ can then be minimized through gradient-based optimization, but there are some difficulties due to eigenvalue degeneracy.
We use a generalized version of degenerate perturbation theory~\cite{Seyranian.1994, johnson2021} to compute a ``generalized-gradient'' that gives a well-defined descent direction even if eigenvalues become degenerate, which makes the regular gradient ill-defined.
Moreover, to allow for meaningful comparison between our initial, uniform parameters and the optimized ones, we impose constraints that keep the final equilibrium yield and total particle concentration fixed at the initial value.
See SI for further details on the minimization of $\tauc$.

Finally, we simulate the dynamical system, Eq.~\eqref{eq:ode}, using the optimized parameters to obtain the optimized yield equilibration time.

\subsection{Optimized parameters improve equilibration times by orders of magnitude}
The result of the equilibration time minimization is shown in the solid lines in Fig.~\ref{fig1}(d).
The difference in equilibration time is drastic: optimizing particle concentrations and binding energies speeds up assembly by over three orders of magnitude, while leaving the final yield nearly identical\footnote{The small differences in the final yield are due to the finite step size used during numerical optimization}.
This speedup is achieved by completely removing the stagnation plateau that was caused by the kinetic trap.
In the optimized system, the yield of the four-particle square never surpasses $5\%$ and the system always maintains a sizable pool of free blue monomers.

To illustrate these results further, in Fig.~\ref{fig2}(a-c), we compare the assembly yield and the required assembly time for three different structures.
Figure~\ref{fig2}(a) shows the tradeoff between assembly yield and equilibration time $\tau_{97}$, here defined to be the first time the target yield reaches $97\%$ of its equilibrium value.
The figure shows that our optimization procedure leads to a speedup of multiple orders of magnitude throughout a wide range of target yields, with the largest speedups often attained at the highest yields.

The benefits of the optimization procedure depend on the system being optimized. 
As an additional example, Fig.~\ref{fig2}(b) shows a semiaddressable 14-particle structure made from triangular building blocks, for which optimization (optimizing 11 particle concentrations and 13 bond energies; 24 parameters in total) results in assembly speedups of over seven orders of magnitude.
In contrast, for the structure shown in Fig.~\ref{fig2}(c), our optimization procedure only leads to speedups of less than a factor of four.

Unfortunately, it is difficult to predict a priori how much assembly can be sped up, but structure topology and the number of possible off-target structures can sometimes be good indicators.
The structure shown in Fig.~\ref{fig2}(c) is a tree, and all off-target structures are also trees where, due to bond promiscuity, one or more ``leafs'' are bound at the wrong place (the blue particle species in particular can (mis)bind in many places).
Little to nothing can be done to prevent those misbound structures from forming, which is why our optimization method does not lead to large speedups.
On the other hand, the structures in Figs.~\ref{fig2}(a-b) are more connected (``bulky''), which generally means that there are many kinetic pathways into the target structure, so that our optimization procedure can identify and optimize the most favorable pathways for fast assembly.

Finally, Fig.~\ref{fig2}(d) shows the relationship between the correlation time (which we have been optimizing for) and the equilibration time of the different systems.
It can be seen that the two times are strongly correlated, which justifies our optimization procedure.

\subsection{Lower addressability can lead to faster assembly}

\begin{figure*}
    \centering
    \includegraphics[width=\textwidth]{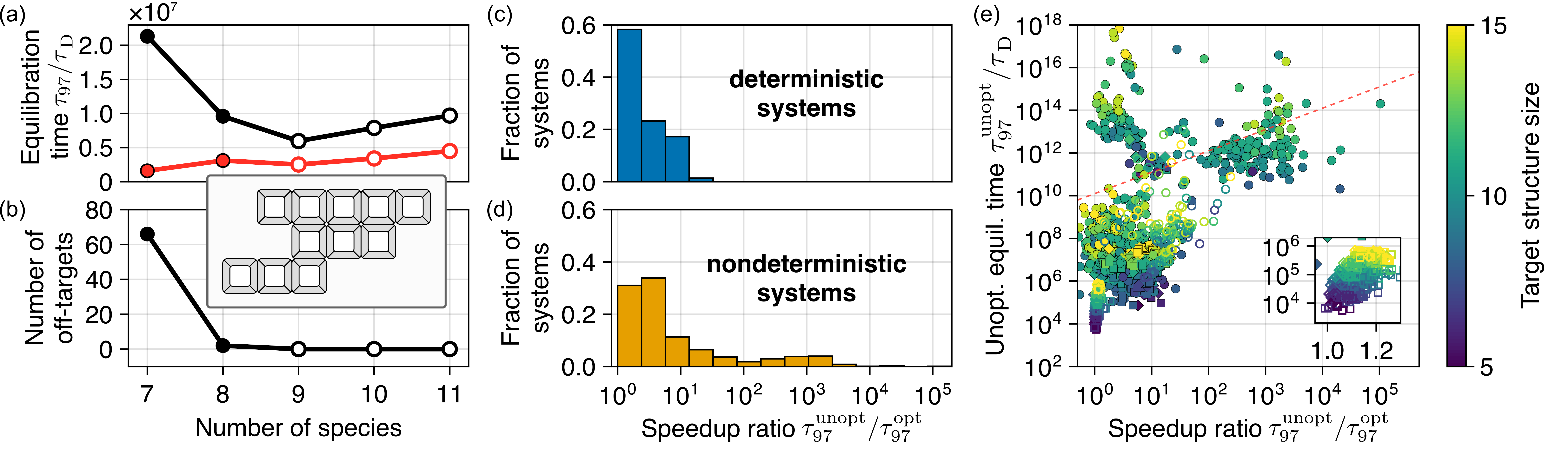}
    \caption{Addressability and equilibration time. (a) Favorable example showing the equilibration time $\tau_{97}$ as a function of the number of distinct particle species for different designs of the same target shape. At all data points, the final yield of the target structure is 90\% and the total particle concentration is $0.1 / \sigma^3$. Unoptimized parameters (black curve) have uniform binding energies across all programmable bonds and stoichiometric concentrations, optimized parameters (red curve) are found using the approach described in Section~\ref{sec:minimize}. (b) The number of off-target structures for different designs of the same target shape. Here, off-target structures are defined as any possible arrangement of particles that is not a substructure of the target. Empty symbols indicate deterministic designs, where the number of off-targets is zero. (c) Histogram of the assembly speedup ratio for 2932 deterministic systems. (d) Histogram of the assembly speedup ratio for 1137 nondeterministic systems. (e) Scatter plot showing the relationship between speedup ratio and the initial, unoptimized equilibration time. Each marker corresponds to one design. Marker color indicates the size (number of particles) in the target structure. Diamond markers correspond to fully addressable designs, square markers to semiaddressable and deterministic designs, circular markers to semiaddressable and nondeterministic designs. Empty markers correspond to systems where the target structure is a tree. Using optimized interactions, systems below the red dashed line equilibrate in less than one day using particles of $20 \, \mathrm{nm}$ radius (see main text).}
    \label{fig3}
\end{figure*}

Figure~\ref{fig2} demonstrates that equilibration times can be lowered, sometimes dramatically, without compromising yield. We now look to address the generality of these results by first considering different binding rules, with different numbers of particle species, for the same target shape, and then analyzing an ensemble of target shapes. 

Figure~\ref{fig3}(a-b) shows an arbitrary target shape made up of 11 square building blocks which displays fairly typical behavior. Starting from the fully-addressable design, where binding rules connect 11 species according to the neighbors in the target structure, we iteratively reduce the number of
species by increasing bond promiscuity, such that some
of the species can fit in multiple locations in the target structure. For convenience, we reject designs that lead to more than 500 possible structures, as well as designs in which the target shape is not ``asymptotically designable''~\cite{Huebl.2025b} (meaning that the equilibrium yield of the target cannot reach 100\% no matter how the design parameters are chosen). This procedure is explained in more detail in Ref.~\cite{Huebl.2025} and in the SI. The result is a series of designs with varying numbers of species that can all assemble the target shape at high yield. 

The equilibration time of these designs is shown in Fig.~\ref{fig3}(a). The black curve shows the unoptimized designs, with stoichiometric concentrations and uniform binding energies set to achieve an equilibrium yield of exactly 90\%. Note that the unoptimized equilibration time slightly \emph{decreases} as the number of particle species is reduced, as long as this lowering of addressability does not lead to any off-target structures, i.e. as long as the assembly process remains deterministic. Figure~\ref{fig3}(b) shows the number of off-target structures, with open circles indicating the deterministic designs for which this is zero.
Reducing the number of distinct particle species increases the configurational entropy of the target structure~\cite{Huebl.2025}, and if a particle can fit in multiple places, it has a higher attachment rate, thus leading to faster assembly.
However, the unoptimized equilibration time increases for nondeterministic systems as the number of off-target structures increases. 

As shown by the red curve in Fig.~\ref{fig3}(a), our optimization procedure, again with the yield fixed to 90\%, causes the equilibration time to decrease significantly for all designs. Interestingly, we do not observe the sharp increase in equilibration times for non-deterministic systems, implying that optimization has a greater effect in the presence of many different structures and thus many assembly pathways. In fact, we are able to achieve a smaller equilibration time for the 7-species design than for the fully-addressable, 11-species design, despite there being 66 off-target structures. 

To systematically quantify these results, we randomly generate 1000 different target shapes, each made up of between 5 and 15 square building blocks, and then generate different binding rules using the procedure described above, generating 3993 systems in total. Figure~\ref{fig3}(c-d) shows the distribution of the speedup ratio, \textit{i.e.} the ratio between the unoptimized and optimized equilibration times, separated according to whether or not the binding rules are deterministic. For deterministic designs, where there are no off-target structures, we are able to achieve a nontrivial speedup, sometimes by as much as $1000\%$. However, speedup ratios in nondeterministic systems can be significantly higher, approaching $10^4$. Figure~\ref{fig3}(e) shows how the speedup ratio correlates with the total equilibration time of the unoptimized designs.
Although far from a clear trend, the data suggest that systems that take longer to assemble also tend to have a higher speedup ratio. In other words, our optimization procedure seems to be most effective where it is needed the most. 

To put these results in perspective, we can estimate the monomer diffusion time as $\taud \approx \eta\, \sigma^3 / (k_\mathrm{B}T)$, where $\eta$ is solvent viscosity and $\sigma$ is the particle size. 
Assuming a particle size of $20\,\mathrm{nm}$, which is typical for DNA-origami-based particles~\cite{Hayakawa.2022pqs, Sigl.2021, Hayakawa.2024, Videbaek.2024}, and assembly at $300\,\mathrm{K}$ in water, this gives $\taud \approx 8\,\mathrm{\mu s}$.
Using this, the red dashed line in Fig.~\ref{fig3}(e) indicates when equilibration, using optimized interactions, is predicted to take a day.
However, note that since the diffusion time scales with $\sigma^3$, this threshold is strongly size dependent.

As already hinted at previously, the most clearly identifiable predictor of speedup is structure topology: we find that the assembly of tree-like structures (i.e. structures with no cycles) can only rarely be sped up dramatically.
This can be seen in Fig.~\ref{fig3}(e) and its inset, which shows a zoomed-in view around the plot origin.
We find that $89.6\%$ of the 1261 tree systems, both fully- and semiaddressable, fall within this region, all of which exhibiting speedups of less than 30\%.
However, beyond the binary classification into trees and non-trees, there is no clear correlation between the speedup ratio, the size, or the number of bonds or ``bulkiness'' of a structure, as shown in Fig.S2 in SI.

\section{Discussion}
We have shown how formulating a self-assembly process as a complex reaction network enables us to predict and design both the assembly outcome and the assembly kinetics.
Optimizing over the binding energies and particle concentrations makes it possible to dramatically speed up the assembly, sometimes by many orders of magnitude, without compromising equilibrium yield.
The potential for speedup is particularly high in semiaddressable systems, which can in some cases even assemble faster than their fully addressable counterparts.
This demonstrates that assembly speed, quality, and economy do not always have to be mutually exclusive -- careful design of binding rules (the primary design space in programmable assembly) and optimization over the binding energies and particle concentrations (the secondary design space) can minimize, and sometimes completely alleviate, the tradeoff between them.

In many cases, the optimized solutions lead to fast assembly by suppressing the (transient) formation of kinetic traps.
We generally observe that our optimization procedure increases the energies of bonds that are formed in substructures that can grow into the desired target, and decreases the energies of other bonds.
In this way, the target structure can be energetically stabilized while at the same time destabilizing the off-targets.
It would be interesting to see if identifying the optimal high-energy and low-energy bonds is possible a priori, without resorting to numerical optimization, by following some design rule that takes into account the target and off-target structures.
With many possible structures and high bond promiscuity, this quickly becomes a daunting combinatorial problem, but perhaps the mathematical tools for designing equilibrium assembly introduced in Ref.~\cite{Huebl.2025b} could be helpful for designing kinetically optimized interactions as well.

Recent work of Jhaveri et al.~\cite{Jhaveri.2024} used a similar reaction-network description to investigate kinetic arrest in fully-addressable assemblies, which, in the very high yield regime, is often caused by monomer starvation~\cite{Hagan.2011, Murugan.2015}.
In contrast, we focused mostly on semiaddressable systems where misbound off-target states are the most dominant kinetic bottleneck; however, we also find that our optimization procedure can lead to speedups even when no off-target structures are possible (Fig.~\ref{fig3}(c)).
Importantly, we operate on a related but different design space: whereas the biologically inspired work of Jhaveri et al. tunes the reaction rates between structures directly, we only tune binding energies and particle concentrations, which are more accessible control knobs in many diffusion-limited synthetic self-assembly systems but offer fewer degrees of freedom with more constraints.
A similar design space was considered by Trubiano et al.~\cite{Trubiano.2021}, who optimized folding pathways of colloidal polymers and also found significant potential for assembly speedups.

Our optimization tools are general and can be applied to a wide range of systems relevant to biology and nanotechnology.
The main requirement is an enumeration of structures that can form during the assembly process.
While obtaining a complete enumeration is generally only possible in systems leading to structures containing less than a few dozen particles, in the future, developing approximate methods like unbiased sampling techniques, or exploiting a separation of timescales in the kinetics, will allow a treatment of larger and larger problems, with increasing experimental applications.
Our tools can be used as a testing ground for potential experimental designs, and offer a fast and easy way to predict and control both assembly outcomes and kinetics in programmable self-assembly.

\acknowledgments
The research was supported by the Gesellschaft für Forschungsförderung Niederösterreich under project FTI23-G-011.

%

\end{document}


\preprint{APS/123-QED}

\title{Supplementary Information: Simultaneous optimization of assembly time and yield in programmable self-assembly}

\author{Maximilian C. H\"ubl}
\author{Carl P. Goodrich}%
\affiliation{%
 Institute of Science and Technology Austria (ISTA), Am Campus 1, 3400 Klosterneuburg, Austria
}%

\maketitle

\section{Enumeration of structures and reactions}
\subsection{Enumeration of structures}
We use the enumeration algorithm introduced in Ref.~\cite{Huebl.2025} to enumerate all structures that are compatible with a given set of binding rules.
This algorithm works by representing structures as directed graphs, and then iteratively builds larger structures from smaller ones by adding individual particles.
Applying graph isomorphism tools to the graph representation makes it possible to avoid generating the same structure multiple times.
All further details can be found in Ref.~\cite{Huebl.2025}.
\subsection{Enumeration of reactions}
Once we have a list of all allowed structures, we can enumerate all possible reactions between structures.
We define a forward/backward reaction reaction pair to be a triple of structures, $a + b \rightleftharpoons c$, where structures $a$ and $b$ can bind to each other to form structure $c$ and structure $c$ can fracture into structures $a$ and $b$.
To find all possible reaction pairs, it suffices to list all possible ways the structures can break into two pieces.

We do this with a basic backtracking search, as follows.
We first represent a structure $c$ as a graph $g_c$ where the graph's nodes correspond to the constituent particles and edges correspond to bonds between particles (note that this representation is different from the one used by the structure enumeration algorithm discussed above, see Ref.~\cite{Huebl.2025} for details.).
We then select an edge $e_1$ of $g_c$, and check if removing that edge disconnects the graph.
If it does, then we have found a possible reaction: we identify the two structures $a$ and $b$ that result from the fracture, store the triple $a + b \rightleftharpoons c$, and move on to the next edge of $g_c$.
If removal of $e_1$ does not lead to fracture, we pick another edge, $e_2$, from the graph, and again check whether removing both $e_1$ and $e_2$ at the same time leads to a disconnect.
If it does, we again store the reaction, otherwise we keep adding edges until either a fracture into two pieces is identified or all edges are exhausted.
We sort the edges lexicographically to avoid generating the same reaction more than once.

\section{Model for aggregation rates}
\subsection{Calculating diffusion coefficients}\label{sec:diffusion}
To compute the aggregation rates corresponding to the binding of two structures, we first need to compute the structures' translational and rotational diffusion constants, or more generally the $6\times6$ diffusion tensor $\mathcal{D}$, describing the combined diffusion of position and orientation of the structure.
We compute $\mathcal{D}$ through an established method that approximates an arbitrarily shaped object as a rigid array of spherical beads~\cite{Harvey.1980, Carrasco.1999, Kraft.2013}.
Since the hydrodynamic friction tensor and the diffusion tensor are related through the Einstein relation, we can use basic hydrodynamics to calculate $\mathcal{D}$.
We use the approach described in Refs.~\cite{Harvey.1980, Carrasco.1999}, where the hydrodynamic friction tensor is calculated by summing over the hydrodynamic interactions and drag forces of all beads when they are put in linear and rotating flows.

Importantly, $\mathcal{D}$ depends on the coordinate origin, and there exists a unique point, called the \emph{center of diffusion}, at which $\mathcal{D}$ is symmetric.
The diffusion constant should be measured from the center of diffusion; otherwise, the translational diffusion constant will be overestimated~\cite{Carrasco.1999}.
Note that the center of diffusion does in general not coincide with the center of mass.

From the eigenvalues of $\mathcal{D}$, measured from the center of diffusion, we can extract an averaged translational and rotational diffusion constant.
As shown in Ref.~\cite{Carrasco.1999}, the diffusion tensor can be written as
\begin{equation}
    \mathcal{D} = \left(\begin{array}{cc}
        \mathcal{D}_\mathrm{tt} & \mathcal{D}_\mathrm{tr}^T \\
        \mathcal{D}_\mathrm{tr}  & \mathcal{D}_\mathrm{rr}
    \end{array}\right) \,,
\end{equation}
where the $3\times 3$ subblocks correspond to translation, rotation, and translation-rotation coupling.
We extract averaged translational and rotational diffusion coefficients from the traces of the subblocks:
\begin{equation}
    D^\mathrm{t} = \frac{\mathrm{tr}(\mathcal{D}_\mathrm{tt})}{3} \,, \quad\quad D^\mathrm{r} = \frac{\mathrm{tr}(\mathcal{D}_\mathrm{rr})}{3} \,.
\end{equation}

The diffusion constants calculated in this way naturally depend on the precise particle geometry, and the results can be made more and more accurate by increasing the ``resolution'', \emph{i.e.}, the number of model beads.
For simplicity, and because we are mainly interested in general results and not in the effects of specific particle geometries, we model all monomers as single spheres of radius $\sigma / 2$.
A structure consisting of $n$ monomers is thus modeled as a rigid array of $n$ spheres.

\subsection{Calculating aggregation rates}
\begin{figure*}
    \centering
    \includegraphics[width=0.8\textwidth]{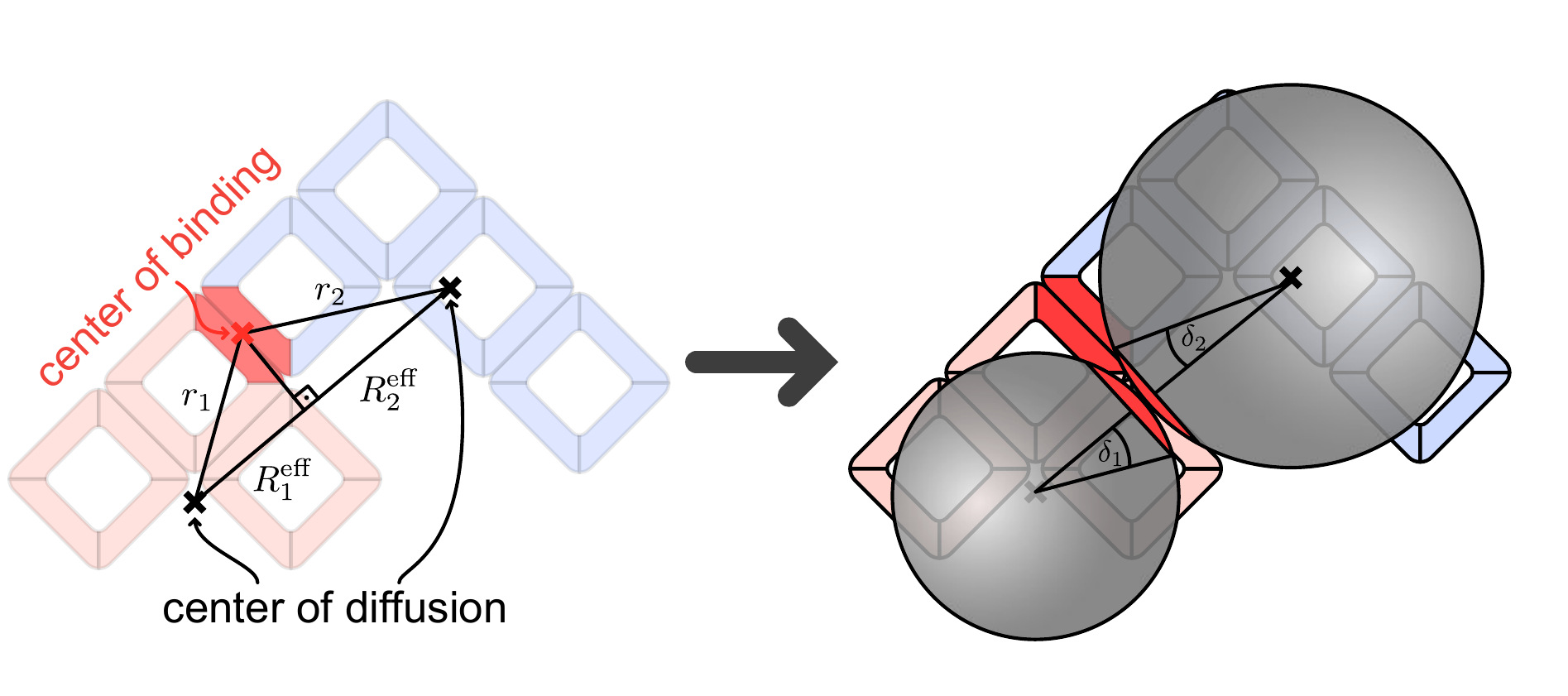}
    \caption{Aggregation rate model. To estimate aggregation rates, we model structures as spheres with an axially symmetric binding patch. The sphere is centered around the center of diffusion of the structure, with radius $R^\mathrm{eff}_i$ determined as defined in the text. The patch angle $\delta_i = \arctan(\sigma / (2R^\mathrm{eff}_i))$ is given by the ratio between $R$ and the characteristic binding site size $\sigma$.}
    \label{fig:ratemodel}
\end{figure*}
We use a coarse-grained model to estimate structure aggregation rates, making use of the results in Refs.~\cite{Solc.1971, Solc.1973, Berg.1985, Zhou.1993, Schreiber.2009}.
As discussed in these references, the binding rate for two spheres of radii $R_1$ and $R_2$, both with an axially symmetric binding patch of opening angle $\delta_1$ and $\delta_2$, is given by
\begin{equation}
    k = 4\pi D R \, \gamma \,,
\end{equation}
where $R = R_1 + R_2$ and $D = D_1^\mathrm{t} + D_2^\mathrm{t}$ is the sum of the particles translational diffusion constants.
The factor $\gamma$ includes the effects of orientational alignment and is given by
\begin{widetext}
\begin{equation}
    \gamma = \frac{F_1F_2}{\Lambda_1\Lambda_2 + [(1 - \Lambda_1)^{-1}(1 - \Lambda_2)^{-1} + (1 - \Lambda_1)^{-1}(\Lambda_2 - F_2)^{-1} + (1 - \Lambda_2)^{-1}(\Lambda_1 - F_1)^{-1}]^{-1}} \,,
\end{equation}
where $F_i = 4\pi\sin^2(\delta_i / 2)$ are the fractional surface areas covered by the patches.
The exact expressions for $\Lambda_i$ involve infinite sums, but it has been shown that they are well approximated by~\cite{Berg.1985, Zhou.1993}
\begin{equation}
    \Lambda_i = F_i\frac{\sqrt{(1 + D^\mathrm{r}_i R^2 / D) / 2} + \cot(\delta_i/2)}{\sqrt{(1 + D^\mathrm{r}_i R^2 / D) / 2} + \sin(\delta_i/2) \cos(\delta_i/2)} \,,
\end{equation}
where $D_i^\mathrm{r}$ are the rotational diffusion constants.
\end{widetext}

To make use of these results, which are valid for axially symmetric spheres, we now construct a mapping that assigns patch angles and radii to every pair of reacting structures, as shown in Fig.~\ref{fig:ratemodel}.
First, we take $R$ to be the distance between the aggregating structures' centers of diffusion.
Then, to compute the patch angles, we assign to each structure an effective radius $R_\mathrm{eff}$ by projecting the distance between the center of the binding area to the center of diffusion onto the displacement vector between the two structures, giving
\begin{align}
    R^\mathrm{eff}_1 = \left| \frac{r_1^2 - r_2^2 + R^2}{2R} \right| \,, \nonumber \\
    R^\mathrm{eff}_2 = \left| \frac{r_2^2 - r_1^2 + R^2}{2R} \right| \,,
\end{align}
where $r_i$ is the distance from the center of structure $i$ to the center of the binding area connected to the other reacting structure.
Note that $ R^\mathrm{eff}_1 +  R^\mathrm{eff}_2 = R$ is generally only true if both structures are convex.
Assuming the binding site area is characterized by a length $\sigma$, we then define the patch angles to be $\delta_i = \arctan({\sigma / (2 R^\mathrm{eff}_i)})$.

\subsection{Equilibrium number densities}
To obtain equilibrium yields and fragmentation rates through detailed balance, we first need to compute the equilibrium number densities.
These can be calculated via structure partition functions, and are given by~\cite{Klein.2018, curatolo.2023, Huebl.2025, Huebl.2025b}
\begin{equation}
    \rhoeq_i = \Omega_i \, e^{\beta(\bm{N}_i\bm{\mu} + \bm{B}_i\bm{\varepsilon})} \,,
\end{equation}
where $\Omega_i$ is the entropic partition function of structure $i$, $\beta = (k_\mathrm{B}T)^{-1}$ is the inverse temperature, $\bm{N}_i$ and $\bm{B}_i$ are vectors listing how many particles of each species and bonds of each bond type, respectively, are part of structure $i$, $\bm{\mu}$ is a vector of chemical potentials of each particle species, and $\bm{\varepsilon}$ is a vector of each binding energy for each bond type.

Following Jhaveri et. al~\cite{Jhaveri.2024}, we neglect cooperativity and thus approximate the entropic partition function $\Omega_i$ as 
\begin{equation}
    \Omega_i = \frac{8\pi^2}{\sigma^3 \chi_i} \,,
\end{equation}
where $\sigma$ is the particle length scale and $\chi_i$ is the symmetry number of structure $i$, denoting the number of rigid body rotations that are equivalent to a permutation of particles. This approximation neglects cooperative effects and is exact for tree-like structures. Such cooperative effects can be included, for example, through the methods discussed in Refs.~\cite{Huebl.2025, curatolo.2023, Holmes-Cerfon.2016, Klein.2018}, but in no way affect the qualitative results of our paper. 

\section{Minimizing correlation times}
To minimize the correlation time $\tauc$, we need to minimize (the real part of) the leading nonzero eigenvalue $\lambda_{n_\mathrm{p} + 1}$ of the system's linear stability matrix $S$ as a function of the particle chemical potentials and binding energies, which we collect in a parameter vector $\pars = (\beta \bm{\mu}, \beta \bm{\varepsilon})$.
The linear stability matrix is obtained from Eq.(1) of the main text by perturbing around the equilibrium state, i.e. setting $\rho_i = \rhoeq_i + \delta\rho_i$, and neglecting terms of $O(\delta\rho^2)$ or higher:
\begin{align}
    S_{ij}(\bm{\xi}) = &\sum_{r \in \mathcal{R}} [\delta_{i c_r} - (\delta_{i a_r} + \delta_{i b_r})] \, \times \\
    &[k_{a_r b_r \to c_r} (\rhoeq_{a_r} \delta_{b_r j} + \rhoeq_{b_r} \delta_{a_r j}) - f_{c_r \to a_r b_r} \delta_{c_r j}]  \,.\nonumber
\end{align}

Minimizing $\lambda_{n_\mathrm{p} + 1}$ is complicated by eigenvalue crossing: even if the leading eigenvalue is non-degenerate in the beginning, the minimization process will often push eigenvalues close together, making $\lambda_{n_\mathrm{p} + 1}$ degenerate.
To resolve this, we adopt the procedure described in Refs.~\cite{Seyranian.1994, johnson2021}, which is a generalized version of degenerate perturbation theory as is commonly used in quantum mechanics.
The basic idea is that, while the gradient of a degenerate eigenvalue is ill-defined, it is still possible to identify a steepest-ascent direction, which can be used for optimization.

Given a matrix $S(\pars)$ with $n$-times degenerate eigenvalues $\lambda_i \equiv \lambda$ and associated right and left eigenvectors $\bm{x}_i$ and $\bm{y}_i$, we can write down the ``generalized gradient''
\begin{equation}
    F^{ij}_\alpha = \sum_{k=1}^n M^{-1}_{ik} \, \bm{y}^T_k \frac{\partial S}{\partial \xi_\alpha} \bm{x}_j \,,
\end{equation}
where $M_{ij} = \bm{y}_i \cdot \bm{x}_j$.
Going along a direction $\Delta\xi_\alpha$ in parameter space, the first-order changes of the degenerate eigenvalues, $\Delta\lambda_i$, are then given by the solutions to
\begin{equation}
    \det\left[\sum_\alpha \Delta\xi_\alpha \, F^{ij}_\alpha - \delta_{ij} \, \Delta\lambda\right] = 0 \,.
\end{equation}

To minimize a degenerate eigenvalue, we need to push all degenerate eigenvalues in the same direction~\cite{johnson2021}, in other words, all $\Delta \lambda_i$ should assume the same value.
One way to achieve this is to find an ascent direction $\Delta\xi^\star_\alpha$ such that
\begin{equation}
    \sum_\alpha \Delta\xi^\star_\alpha \, F^{ij}_\alpha = \delta_{ij} \,.
\end{equation}
It can be shown that this linear system is always non-singular, except when optimality is reached~\cite{Seyranian.1994}, meaning that this method always leads to a meaningful ascent direction.

In Ref.~\cite{Seyranian.1994}, it is further shown how to impose additional constraints on $\Delta\xi^\star_\alpha$ by adding additional linear equations to the system above.
This lets us impose a constraint of constant yield and constant particle concentration during the optimization, so that we can find the maximal assembly speed-up without changing  yield or the total particle concentration.

\section{Equilibration time and structure bulkiness}
\begin{figure}
    \centering
    \includegraphics[width=\linewidth]{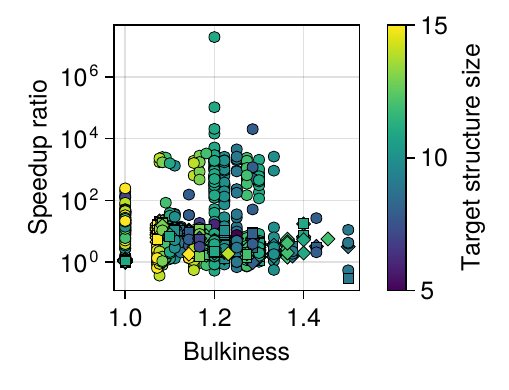}
    \caption{The relationship between structure bulkiness (see text) and the speedup ratio, as defined in the main text. Marker color indicates the size (number of particles) in the target structure. Diamond markers correspond to fully addressable designs, square markers to semiaddressable and deterministic designs, circular markers to semiaddressable and nondeterministic designs. The underlying systems are the same as the ones treated in Fig.3(e) of the main text.}
    \label{fig:bulk}
\end{figure}

As discussed in the main text, we find that tree-like structures can only rarely be sped up dramatically.
To investigate the effect of structure topology further, we quantify the ``bulkiness'' of a structure as
\begin{equation}
    b = \frac{n_\mathrm{b}}{n_\mathrm{p} - 1} \,,
\end{equation}
where $n_\mathrm{b}$ is the number of bonds in a structure and $n_\mathrm{p}$ is the number of particles in a structure.
Tree-like structures always have a bulkiness of $b=1$, whereas more fully connected structures have a bulkiness of $b > 1$.
Fig.~\ref{fig:bulk} shows the relationship between the speedup ratio (as defined in the main text) to this measure of bulkiness.

\section{Generating assembly designs}
Here we explain the generation of structures and binding rules used to generate the data for Fig.~3 of the main text.

We first generate 1000 shapes made from squares by applying the structure enumeration tools of Ref.~\cite{Huebl.2025}, but with one additional step:
During the enumeration process, we randomly reject some of the structures.
More precisely, for each structure with a given size (number of constituent particles) $n_\mathrm{p}$ and bulkiness $b$ (see above), we accept this structure with probability
\begin{equation}
    p_\mathrm{accept} = 
    \begin{cases}
        0.95^{n_\mathrm{p}} \, \quad &\text{if } b < 1.5\, \\
        1  \, \quad &\text{else} \,.
    \end{cases}
\end{equation}
This stochastic enumeration process allows us to (1) quickly generate structures of larger sizes, and (2) generate a more even mix between tree-like shapes and bulky shapes: completely random generation leads to a very high percentage of tree-like structures, and we choose this form of the acceptance probability to counteract this.
We repeatedly run structure enumerations with this stochastic rejection rule until we generate more than 1000 structures total, and from those, then randomly select exactly 1000 structures to be used going forward.

For each target shape, we then generate different sets of binding rules that lead to the assembly of the target.
Starting from the fully-addressable design corresponding to each target (where every particle in the target is its own, individually addressable species), we iteratively reduce the number of species by increasing bond promiscuity, such that some of the species can fit in multiple locations in the target structure.
This procedure is explained in detail in Ref.~\cite{Huebl.2025}.
To speed up the generation of binding rules, we do not iterate through all possible ways particle species can be reduced, but simply pick the binding rules that generate the fewest structures for a given number of species, and then use these rules as the starting point from which the next particle species is eliminated.
As mentioned in the main text, we reject designs that lead to more than 500 possible structures, as well as designs in which the target shape is not ``asymptotically designable''~\cite{Huebl.2025b}.
The end result is a large collection of different binding rules, each capable of assembling one of the example targets.

To evaluate the speedup of each set of binding rules, we first choose uniform binding energies and stoichiometric particle concentrations, such that the yield of the target structure is exactly $90\%$.
We then optimize the equilibration time of each system using the procedure described in the main text, and measure the ratio between the unoptimized and optimized equilibration times, which we call the ``speedup ratio.''
Note that this optimization process did not converge for around 2\% of the initially generated systems, since in some cases, manual tuning of optimization hyper-parameters would be necessary to achieve convergence.
We exclude systems for which the optimization did not converge without manual intervention, leading to a final total of 3993 binding rules, each capable of assembling one of the 1000 target shapes.

%